\documentclass[aps,prb,preprint,superscriptaddress]{revtex4}
\usepackage{graphicx}
\usepackage{mathptm}
\begin{document} 
\draft
\title{Multiple scattering and attenuation corrections in
Deep Inelastic Neutron Scattering experiments} 

\author{J. Dawidowski, J.J. Blostein and J.R. Granada} \affiliation{Consejo Nacional de
  Investigaciones Cient\'{\i}ficas y T\'ecnicas, Centro At\'omico Bariloche and
  Instituto Balseiro, Comisi\'on Nacional de Energ\'{\i}a At\'omica,
  Universidad Nacional de Cuyo,(8400) Bariloche, Argentina}
\email{javier@cab.cnea.gov.ar}

\date{\today}
\begin{abstract}
  Multiple scattering and attenuation corrections in Deep Inelastic
  Neutron Scattering experiments are analyzed. The theoretical basis
  is stated, and a Monte Carlo procedure to perform the calculation is
  presented. The results are compared with experimental data. The
  importance of the accuracy in the description of the experimental
  parameters is tested, and the implications of the present results on
  the data analysis procedures is examined.
    
\end{abstract}


\maketitle

\section{Introduction} 
\label{intro} 
Since its creation in 1966 by Hohenberg and Platzmann \cite{Hohe}, the
Deep Inelastic Neutron Scattering (DINS) technique was considered the
most direct probe of the momentum distribution of nuclei in condensed
matter. The interest on this technique was further stimulated by
subsequent developments, that showed that features attributed to the
interference between the neutron and the proton could be observed in
experiments made on hydrogen \cite{May1}, thus determining the wave
function of the protons directly from the experimental data.  The
availability of this technique as a customary research tool, opened in
the last decade a new field for the investigation in condensed matter,
and stirred up particular interest in the study of hydrogen dynamics,
a topic for which this technique is particularly suitable.  Despite
the main activity in the field is held at Rutherford Appleton
Laboratory (United Kingdom), contributions from different laboratories
were also reported in the past \cite{LosAlamos,LosAlamos2,Japon} and
recently \cite{Wang,Dawn2m,Blosmun,Blosjap}.
 
The technique basically consists in an energy analysis carried out
through the use of neutron resonant filters in the range of a few
electron-Volts, and it is based on pulsed neutron sources.  The
spectra are recorded in time channels, which for the purpose of the
study of momentum distributions must be translated to a momentum
variable, thus obtaining the so-called Neutron Compton Profile (NCP).
 
The theoretical basis of the technique was established by Sears
\cite{Sea1}, who outlined the general procedure that must be employed
to arrive to the NCP from the experimental data. Later, Mayers
\cite{May2} established the conditions of validity of the usually
employed approximations.
 
Recently, we re-examined the usual procedure to analyze the
experimental data obtained from this technique \cite{Blos,Blos2}, and
and suggested improvements in the analysis of experiments involving
light nuclei \cite{PRB}.  In the cited reference, we showed that the
method to obtain the momentum distributions is, in general, a
non-trivial task that involves a thorough knowledge of the different
components of the experimental setup.  In particular, it is important
to know the energy spectrum of the incident neutrons, the detector
efficiency, as well as an accurate description of the filter total
cross section.  However, in the above mentioned work we did not
examine the sample-dependent effects such as multiple scattering and
attenuation, which must be previously accounted for, before any
analysis is attempted. It is a very well-known fact that multiple
scattering and attenuation effects can be important even if all the
reasonable cautions are taken in the sample design, since some
low-signal portion of the observed spectra could nevertheless be
seriously affected \cite{Daw1}.
 
Multiple scattering corrections is a long-debated subject, and has
been extensively treated in the literature.  Following the pioneering
works of Vineyard \cite{Vine}, and Blech and Averbach \cite{Blec},
Sears thoroughly established its theoretical basis \cite{Sea2},
stating the integrals that describe the contribution of the n-th
scattering process to the observed spectrum. The complexity of its
solution was successfully tackled by Copley \cite{Copl} who devised a
Monte Carlo code, suited to a particular experimental situation.
However, Copley's scheme could not be easily adapted to different
kinds of experiments.  On the other hand, the common knowledge in
neutron scattering technique teaches that specific procedures must be
devised for each particular experimental situation \cite{Daw1,Daw2}.
 
Therefore, there is a primary need of a customary correction tool for
multiple scattering and attenuation effects in DINS experiments, that
was only recently fulfilled. The authors have already presented
numerical results from a new Monte Carlo code compared with
experimental results in multiple scattering corrections in DINS
experiments (see Refs. \cite{Dawn2m,Blosjap,Blosmun}), and recently
Mayers {\it et al.} \cite{MayeMS} introduced the details of a new
Monte Carlo code, related with the experimental setup thoroughly
described in Ref. \cite{Fiel}. The Monte Carlo procedure described in
the mentioned work contains a series of assumptions that were
carefully analyzed in several publications. In the first place it is
considered that the neutron has a well-defined final energy
corresponding to the maximum of the main absorption of the resonant
filter. In a recent analysis \cite{finalE} we showed that such
distributions are far more complex than considering a single final
energy, and depends on the time channel and the dynamics of the
scattering species. The scattering angles in \cite{MayeMS} are
generated considering a random isotropic distribution. The validity of
such assumption was investigated in Ref. \cite{Daw0} showing that the
results obtained from such approximation deviate significantly from
the exact ones when considering incident neutrons in the epithermal
range. Another approximation contained in the mentioned work the
description of the total cross sections of the scattering system with
a constant plus a `1/v' absorption scattering law, which can deviate
significantly from the real behavior when considering molecular
systems \cite{Gra2}.

In this paper we present the fundamentals of our Monte Carlo procedure
to account for multiple scattering and attenuation effects. Account is
taken of the experimental details such as the energy spectrum of the
incident neutrons, the resonant filter transmission and the efficiency
of the detectors bank.  Regarding the sample, inelastic scattering is
taken into account employing suitable models for each analyzed case.
The proposed model can be easily introduced as a double-differential
cross section, either analytical or numerically, so a complete
description of the energy-transfers and scattering angles are obtained.
Experimental results of samples of different sizes are shown, and the
present code is benchmarked. The importance of an accurate model to
describe the neutron-sample interaction is stressed, and finally
implications of multiple scattering corrections on different
situations are discussed.
 
\section{Basic formalism} 
\label{basic} 
 
A basic description of the kind of experiments  that we will analyze can
be found in Ref. \cite{Blos}, so we will give only a 
brief account here. We will analyze the case a DINS experiment 
performed in a pulsed source with an inverse-geometry configuration, 
\textit{i.e.} the resonant filter is placed in the path of the 
neutrons emerging from the sample. A typical experiment consists of
alternative 'filter out' and 'filter in' measurements, whose  difference
gives the NCP in the time-of-flight scale. 
 
We define $E_0$ as the incident neutron energy (characterized by a
spectrum $\Phi(E_0)$), $E$ its final energy, $L_0$ the source-sample
distance, $L_s$ the sample-detector distance and $\theta$ the scattering
angle. The total time elapsed since the neutron is emitted from the
source until it is detected, for a single-scattering event, is
\begin{equation} 
t=\sqrt{\frac{m}{2}}\Bigl(\frac{L_0}{\sqrt{E_0}}+\frac{L_s}{\sqrt{E}}\Bigr) , 
\label{tof} 
\end{equation} 
where $m$ is the neutron mass. 
 
The resonant filter will be characterized by a total cross section
$\sigma_F(E)$, a number density $n$ and a thickness $T$, so the fraction
of neutrons transmitted by it will be $\exp{(-nT\sigma_F(E))}$.  If
$\displaystyle\frac{d^2\sigma}{dEd\Omega} (E_0,E,\theta)$ is the sample
double-differential cross section, then the difference count rate
('filter-out' minus 'filter-in') at time of flight $t$, will be
\cite{Blos,Powl}
\begin{equation} 
c(t) = \mathop{\int_{\scriptstyle {\rm E_{inf}}}^{\rm \infty}}_{\rm t=const}dE_0\,\Phi(E_0) 
\frac{d^2\sigma}{d\Omega dE}(E_0,E,\theta)\,\varepsilon(E)  
\,(1-e^{-nT\sigma_F(E)}) 
\Big|\frac{\partial E}{\partial t}\Big|\Delta\Omega,
\label{ct} 
\end{equation} 
where $\varepsilon(E)$ is the detector efficiency, and $\Delta\Omega$ the solid angle
subtended by the detector.

Integral (\ref{ct}) must be calculated at constant time, taking into
account relationship (\ref{tof}) between $E_0$ and $E$, and the
Jacobian $\big|\partial E/\partial t\big|$ must be evaluated at a fixed time
\cite{Powl}.  The lower limit of integration is determined by the
condition that in the second flight path the neutron has infinite
velocity, \textit{i.e.}  $E_{\rm inf}=\frac{1}{2}m L_0^2/t^2$. It is worth
remarking that Eq. (\ref{ct}) is a valid expression only if single
scattering events would take place.
 
\section{Multiple Scattering} 
\label{ms} 
 
In this section we will give an outline of the basic equations 
which govern the multiple scattering processes of n-th order, 
that are employed in the Monte Carlo programs. For a more detailed 
treatment the reader is referred to  \cite{Daw1} and \cite{Sea2}. 
 
We will suppose throughout this paper an isotropic sample. 
Let $S(Q,\omega)$ be the scattering law of the sample, $E_0$ and 
$E$ the incident and final neutron energies (being $\mathbf{k_0}$ 
and $\mathbf{k}$ their corresponding wave vectors), $d\Omega$ the 
element of solid angle in the direction of the scattered neutron, 
$\sigma_b$ the bound-atom scattering cross section of the 
sample (considered monatomic), and $N$ the number of scattering
centers, then the double-differential cross section is 
\begin{equation} 
\frac {d^2 \sigma}{d\Omega dE}= \frac {N\sigma_b}{4\pi} \frac {k}{k_0} 
S(Q,\omega), 
\label{smicro} 
\end{equation} 
defined as the average number of scattered neutrons with final 
energies between $E$ and $E+dE$, and within a solid angle 
$d\Omega $, per unit incident flux. As usual we define  
${\bf Q}={\bf k}_0-{\bf k}$ and $\hbar\omega=E_0-E$. 
The integral of Eq. (\ref{smicro}) over all angles and final energies 
gives the microscopic total cross section $\sigma(E_0)$. 
 
The above definition corresponds to the ideal textbook case 
where there is not multiple scattering. Turning to the real 
case let us define the \textit{macroscopic double-differential 
cross section} as the probability that an incident neutron with a 
wave-vector ${\bf k}_0$ \textit{will emerge from the sample} with a 
wave-vector ${\bf k}$ \cite{Sea2}. In this definition we do not take into 
account neutrons non-interacting with the sample (i.e. transmitted). 
Its expression thus reads 
\begin{equation} 
\frac {d^2 \Sigma}{d\Omega dE}=\frac {1}{4\pi A} \frac {k}{k_0} 
s(\mathbf {k}_0,\mathbf {k}), 
\label{smacro} 
\end{equation} 
where $A$ is the cross-sectional area perpendicular to the incident
beam.  $s(\mathbf {k}_0,\mathbf {k})$ is an effective scattering
function that admits a decomposition in a part due to singly-scattered
neutrons in the sample $s_1(\mathbf {k}_0,\mathbf {k})$, another due
to singly-scattered neutrons in the container $s_C(\mathbf
{k}_0,\mathbf {k})$, and a third due to multiply scattered neutrons
(with any combination of sample-container scattering events)
$s_M(\mathbf {k}_0,\mathbf {k})$
\begin{equation} 
s(\mathbf {k}_0,\mathbf {k})=s_1(\mathbf {k}_0,\mathbf {k})+ 
s_M(\mathbf {k}_0,\mathbf {k})+s_C(\mathbf {k}_0,\mathbf {k}). 
\label{decomp} 
\end{equation} 
The single scattering component $s_1$ is simply related with 
the scattering law through the relationship 
\begin{equation} 
s_1(\mathbf {k}_0,\mathbf {k})=N\sigma_bS(Q,\varepsilon)H(\mathbf {k}_0,\mathbf {k}), 
\label{factrans} 
\end{equation} 
where $H(\mathbf {k}_0,\mathbf {k})$ is the first-order attenuation
factor, defined as the fraction of single-scattered neutrons that fail
to leave the sample due to multiple scattering and nuclear absorption
\cite{Sea2} or that are not detected due to the detector efficiency
\cite{Daw1}. Expression (\ref{smacro}) inserted into (\ref{ct}), gives
the real NCP including multiple scattering components.  Its
calculation, will normally involve a numerical simulation based on the
Monte Carlo method.
 
\section{Monte Carlo code} 
 
\label{mc} 
 
In this section we will describe the numerical simulation  devised for
DINS experiments. Its fundamentals are based on  Copley's method
\cite{Copl}, and they are extensively developed in \cite{Bisc} and \cite{Span}. 
 
\subsection{Neutron Histories} 
\label{nh} 
Neutron histories are generated with an initial unity weight.  The
incident neutron energy is decided randomly using the experimental
neutron spectrum as the probability distribution. The flight path $x$
for a neutron with energy $E$ is given by the probability
\begin{equation} 
p(E,x)=\frac{\Sigma_t(E,x){\mathcal T}(E,x)}{1-{\mathcal T}(E,d)}, 
\label{pex} 
\end{equation} 
where the probability has been biased so the neutron never gets 
out of the sample \cite{Bisc}. In Eq. (\ref{pex}), $\Sigma_t(E,x)$ is the 
macroscopic total cross section of the sample-container set 
a distance $x$ away from the neutron previous scattering 
position, taken in the current flight direction, 
${\mathcal T}(E,x)$ is the fraction of noninteracting 
(transmitted) neutrons in that direction after traversing 
a distance $x$, and $d$ is the distance to the sample 
surface in that direction. To compensate the bias in the probability, 
a weight is assigned to each neutron which decreases 
according to the transmitted fraction in the traversed path, being 1 
the initial value. Given the weight at step $i-1$  the weight at step 
$i$ is calculated as  \cite{Span}
\begin{equation} 
w_i=w_{i-1}(1-{\mathcal T}(E,d)) \frac {\Sigma _s(E,0)}{\Sigma _t(E,0)}, 
\label{peso} 
\end{equation} 
where $\Sigma _s(E,0)$ and $\Sigma _t(E,0)$ are the macroscopic scattering and
total cross sections, respectively, at position $i-1$ and its ratio
indicates the probability that the neutron will not be absorbed in the
considered path. A history is finished when the weight drops under a
predetermined cut-off value, so the number of scattering events is not
predetermined.
 
The assignment of new energies and flight directions is made via the
use of model distributions for the double-differential cross sections
of the sample and the container environments, normalized with the
total cross section at the current energy $E_i$ \cite{Daw1}
\begin{equation} 
P(E_i,E,\theta )= \frac {N\sigma_b}{4\pi\sigma(E_i)} \frac {k}{k_i} 
S_{model}(Q,\omega). 
\label{probmod} 
\end{equation}

\subsection{Scoring} 
\label{sc} 
 
At each step, the contribution of the current history to the detectors
is calculated for each time-of-flight channel $t$. The final energy
$E$ to be considered for this channel is obtained from \cite{Daw1}
\begin{equation} 
t=\sqrt{\frac{m}{2}}\left(\frac{L_0}{\sqrt{E_0}}+ 
\sum_{i=1}^{N} \frac{L_i}{\sqrt{E_i}}+\frac{L_s}{\sqrt{E}} \right) 
\label{tofe} 
\end{equation} 
where $N$ is the number of scattering steps, and $L_i$ is the flight 
path of step $i$, which was covered with an energy $E_i$. 
 
The quantity to be scored is the current weight, times the
transmission factor from the current position to the sample surface in
the direction towards the detector position, times the filter
absorption ratio, times the detector efficiency
\begin{equation} 
z_i= w_i P(E_i,E,\theta) {\mathcal T}(E,d) (1-e^{-nT\sigma_F(E)}) \varepsilon(E). 
\label{zsc} 
\end{equation} 
It can be shown \cite{Daw1} that the average of $z_i$ after a large number of 
histories is the sought solution of Eq.(\ref{ct}) for the case 
of a macroscopic sample (Eq.(\ref{smacro})). 

\subsection{Summary of input data}
 
The above described Monte Carlo procedure, requires a detailed
description of the experimental setup and total cross sections of the
involved materials. It also makes use of models for the scattering
laws to describe the sample and container interaction with neutrons.
Here we summarize the input data needed to perform it.
\begin{itemize}
\item  Incident spectrum as a function of energy.

\item  Total cross section of the sample and the container materials
as a function of the energy. These data must be tabulated in an energy
range wide enough to cover not only the energies corresponding to the
incident spectrum (epithermal energies), but also to consider the energy 
transfers after a number multiple scattering steps (typically thermal 
energies).

\item  Mean free path of the sample and the container as a function of energy.

\item  Detector bank efficiency as a function of energy.

\item  Input parameters for the chosen models to describe the 
scattering law of the sample and the container. Alternatively the
models can be defined through a numerical input.

\item  Geometry parameters for the proposed experimental setup and
sample environment.

\item  Total cross section of the resonant filter in an energy range
comprising thermal neutrons (to give a good description of the '1/v'
region \footnotemark), to energies above the main resonance (to describe the lower
time-of-flight channels).
\end{itemize}

\footnotetext {Here we refer to the slow neutron regime, where the
absorption cross section is inversely proportional to the neutron
velocity. See Ref. \cite{Blat}.}

\section{Experimental setup} 

The experiments were performed at the Bariloche pulsed neutron source
(Argentina). Neutrons, produced by the interaction of the electrons 
accelerated by the LINAC on a lead target, are moderated in a 4 cm 
thick polyethylene slab. A cadmium sheet is placed in the incident 
beam, to absorb thermal neutrons. The LINAC was operated at a 100 Hz
rate.  A collimated neutron beam 1 inch diameter was employed.

\begin{figure}
\begin{centering}

\resizebox{0.6\textwidth}{!}{\includegraphics{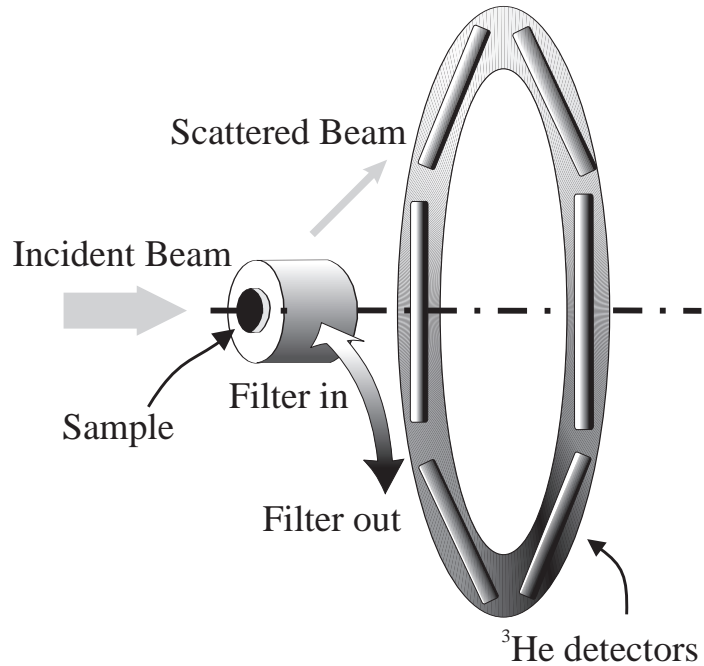}}
\caption{Experimental setup employed for DINS experiments. The detectors
are placed at a scattering angle of 56$^0$.}
\label{setup}
\end{centering}
\end{figure}
A schematic view of the of the DINS facility is shown in Fig.
\ref{setup}. A movable cylindrical indium filter 0.25 mm thick, is
placed in the flight path of the scattered neutrons. The movement is
controlled remotely to perform alternative 'filter-in' and
'filter-out' measurements every 10 minutes. The detector bank consists
of six $^3$He proportional counters (10 atm filling pressure, 6 inch
active length, 1 inch diameter) placed at a mean scattering angle of
56$^0$. The detectors were covered with cadmium cylinders to minimize
the background due to thermal neutrons. The flight-path lengths were
504 cm (source-sample distance), and 27.5 cm (sample-detector)
respectively.
\begin{figure}
\begin{centering}
\resizebox{0.8\textwidth}{!}{\includegraphics{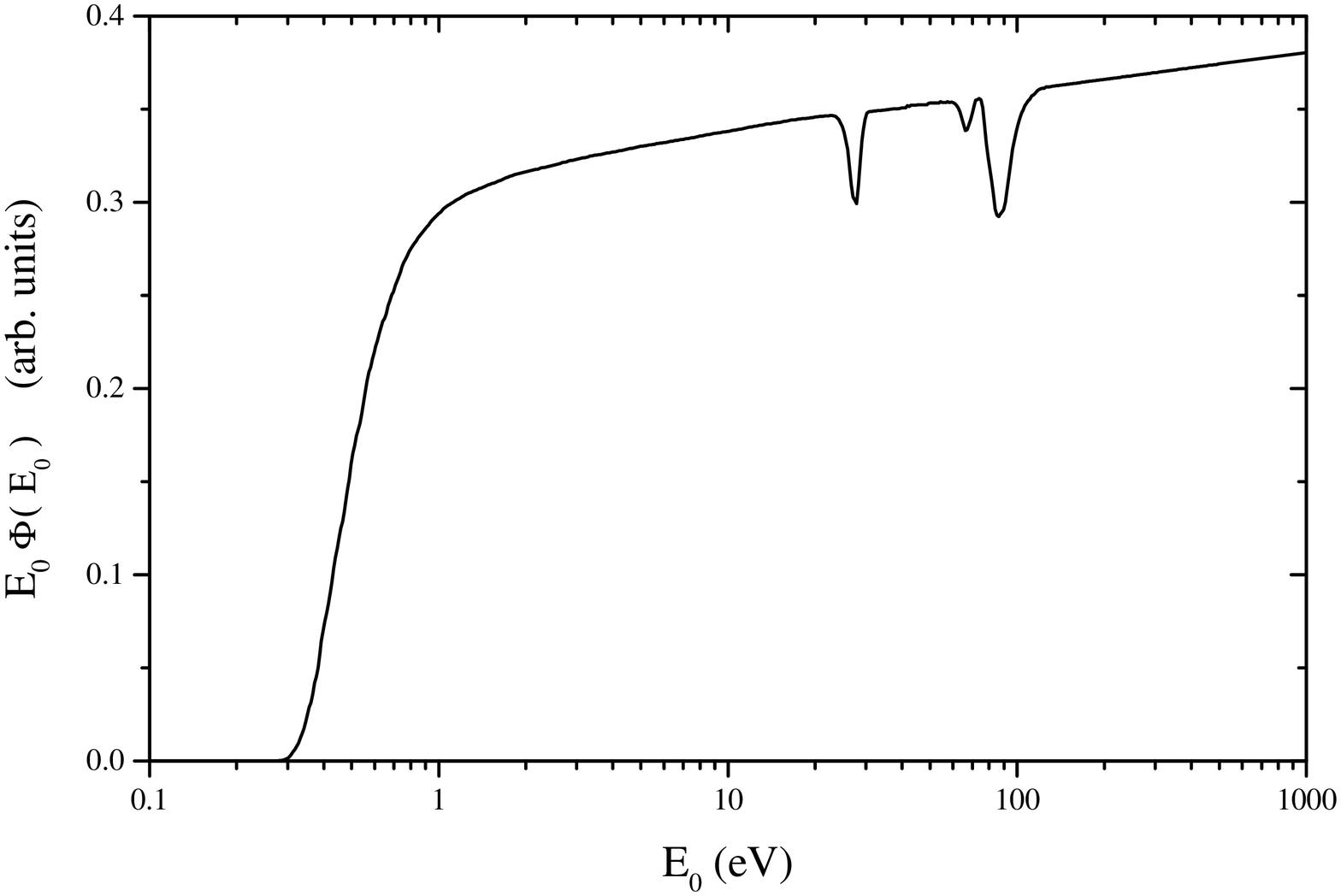}}
\caption{Incident neutron spectrum multiplied by the energy in order
to represent it in logarithmic $E_0$ scale. Dips due to resonances in
the cadmium sheet are observed.}
\label{incid}
\end{centering}
\end{figure}

\begin{figure}
\begin{centering}
\resizebox{0.8\textwidth}{!}{\includegraphics{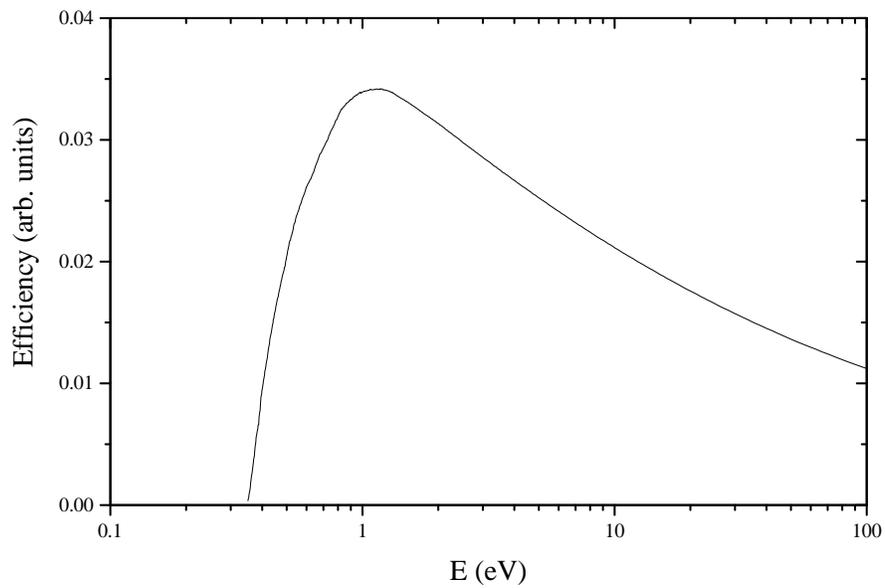}}
\caption{Detector bank efficiency. The cutoff about 0.5 eV is due to the
cadmium cylinders which cover the detectors.}
\label{eff}
\end{centering}
\end{figure}

The incident spectrum, measured employing a $^3$He detector placed
perpendicularly to the direct beam, is shown in Fig. \ref{incid} where
the detector efficiency effect was accounted for. The detector bank
efficiency was determined through the ratio of the spectrum of
scattered neutrons on a lead sample (which is mostly an elastic
scatterer), and the spectrum measured on the direct beam. The result
is shown in Fig.  \ref{eff} where the cutoff near 0.5 eV is due to the
cadmium cylinders that cover the detectors.
 
\section{Results and Discussion} 

In this section we will analyze our experimental and numerical results
on the Compton profiles for coin-shaped graphite and polyethylene
samples of different sizes at room temperature.  In the further
paragraphs we will show results on samples whose sizes were chosen to
serve as a benchmark on the numerical simulations, and they are not
intended to represent optimized choices in the experimental design.
Finally we will show the importance of the present corrections in thin
samples, suitable for the experimentalists' work. Numerical
simulations were performed employing the above mentioned Monte Carlo
code, making use of the experimental parameters mentioned in the
previous section, and using total cross section data for the Indium
filter from Ref. \cite{Mcla}.

\begin{figure}
\begin{centering}
\resizebox{0.8\textwidth}{!}{\includegraphics{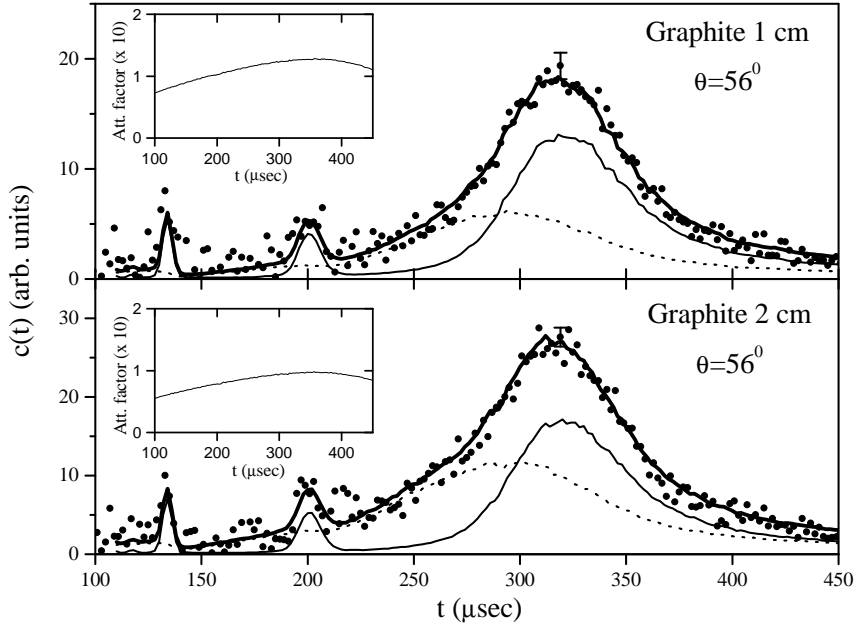}}
\caption{NCP for the two measured graphite samples. Normal line:
single scattering component; dotted line: multiple scattering; thick line:
total scattering. Insets: attenuation factors}
\label{graphite}
\end{centering}
\end{figure}

In Fig. \ref{graphite} we show the results for the graphite samples,
3.54 cm diameter and 1 and 2 cm thickness respectively. A typical
error bar is indicated for each experimental dataset. The approximate
difference in count-rate ('filter out' minus 'filter in') at the peak
maximum was 4 counts every 10000 LINAC pulses for the thin sample and
6 counts every 10000 LINAC pulses for the thick one. Measurements were
carried out in 4 million LINAC pulses for the thin and 2 million LINAC
pulses for the thick sample. Numerical simulations were carried out
using a gas model for the graphite. This is a good approach at
epithermal energies like in this case, with the condition that the
temperature must be replaced by an effective temperature that takes
into account the phonon dynamics \cite{Gra1}. The resulting effective
temperature is 61.2 meV calculated on the basis of a Debye temperature
of 1860 K \cite{Ashc}.  In Fig. \ref{graphite} we show the Monte Carlo
results for the single and multiple scattering components as well as
the total one. Besides the main peak at about 315 $\mu$sec two extra
peaks at 200 and 130 $\mu$sec are observed, due to the resonances of
3.85 and 9.07 eV of Indium, respectively. A good agreement is observed
between the calculation and the experimental data, showing that the
multiple scattering component has a peak shape that is broader than
the main peak and is centered at lower times of flight, thus
contributing to a significant distortion in the observed total
scattering. In the insets of Fig. \ref{graphite} we show the
attenuation factors (Eq.  (\ref{factrans})), that have to be applied
to the observed profile, once multiple scattering is subtracted.

\begin{figure}
\begin{centering}
\resizebox{0.8\textwidth}{!}{\includegraphics{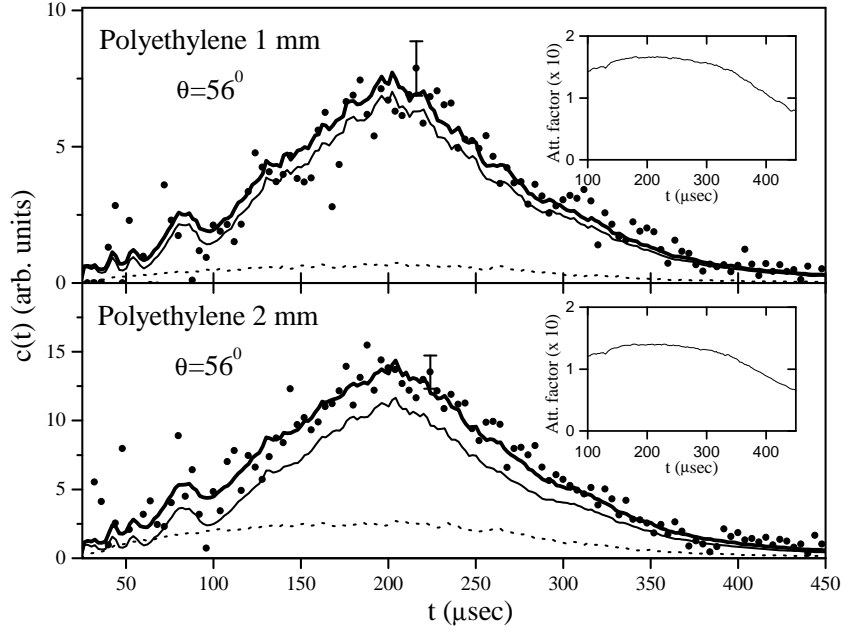}}
\caption{NCP for the two measured polyethylene samples. The same
notation as in Fig. \ref{graphite} applies. Insets: attenuation factors}
\label{poly}
\end{centering}
\end{figure}

In Fig. \ref{poly} we show the same results for the two samples of
polyethylene 3 cm diameter and 1 and 2 mm thickness respectively,
where typical error bars are shown. The approximate count rate at the
peak position was 5 counts every 10000 LINAC pulses for the thin
sample and 7.5 counts every 10000 LINAC pulses for the thick one.
Measurements were carried out in 5 million LINAC pulses for the thin
and 4 million LINAC pulses for the thick sample. For the Monte Carlo
simulations we employed the Synthetic Model \cite{Gra2} with the
parameters from Ref. \cite{Daw3}, which was successfully employed to
describe different integral magnitudes of the double-differential
cross section. The model adequately describes the interaction between
the neutron and the sample in different energy regimes, tending
naturally to the commonly employed impulse approximation in the
epithermal region. Although the shape of the observed main peak due to
hydrogen is less affected by multiple scattering effects than in the
case of graphite, it must be noted the distorting effect due to the
attenuation factor that varies a 25 \% from 200 to 350 $\mu$sec thus
affecting significantly the long-times tail of the Compton profiles.

\begin{figure}
\begin{centering}
\resizebox{0.8\textwidth}{!}{\includegraphics{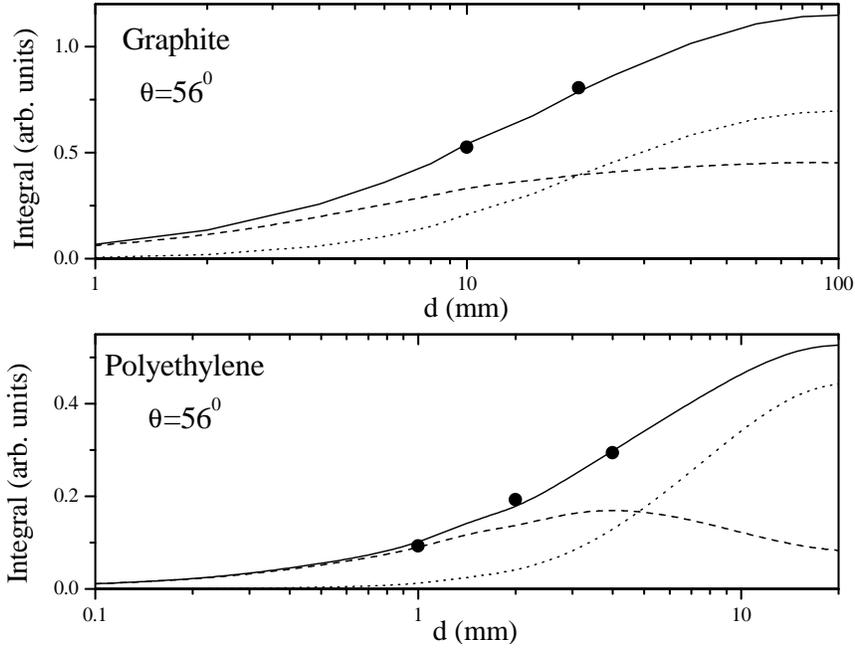}}
\caption{Integrated intensity for single (dashed line), multiple (dotted line) 
and total scattering (full line) for different thicknesses of graphite and 
polyethylene samples, compared with the experimental data (black circles).}
\label{integ}
\end{centering}
\end{figure}

The general trend of the multiple scattering behavior can be analyzed
by observing the total intensities observed in the main peak of the
Compton profiles as a function of the sample thickness. In Fig.
\ref{integ} we show the integral intensity of the main peak for
single, multiple and total scattering obtained from Monte Carlo
simulations at several sample thicknesses of graphite and
polyethylene. In the same graph we show the results obtained from our
experimental data. In the case of polyethylene, we measured a third
sample 4 mm thick, that is not included in Fig. \ref{poly}, but is
shown in Fig. \ref{integ}. It is worth to mention that the results
from the simulations were multiplied by a constant (the same value in
all the cases) in order to fit the experimental data. In both systems,
we observe that the trend of the peak intensity as a function of the
sample thickness can be correctly accounted for only if multiple
scattering processes are considered.

It is worth to emphasize the importance of a good description
of the filter total cross section and the detector efficiency. For that
purpose we performed simulations assuming two different cases:
\begin{enumerate}
\item[(a)] a black detector (\textit{i.e.} a detector with unit efficiency)
and the filter described by the complete absorption cross section;
\item[(b)] a filter described with a Lorentzian shape\footnote{
A Lorentzian centered at 1.457 eV and full-width at half maximum of
0.2016 eV was employed. This value is greater than the one found
in the literature \cite{Mugh} (0.075 eV at 0~K) due to the finite
temperature of the filter. The peak shape at finite temperature is not a
Lorentzian but a peak described with the Lamb equation \cite{Beck} ,
that is broader due to the Doppler effect.}, with the real detector
efficiency.
\end{enumerate}

\begin{figure}
\begin{centering}
\resizebox{0.8\textwidth}{!}{\includegraphics{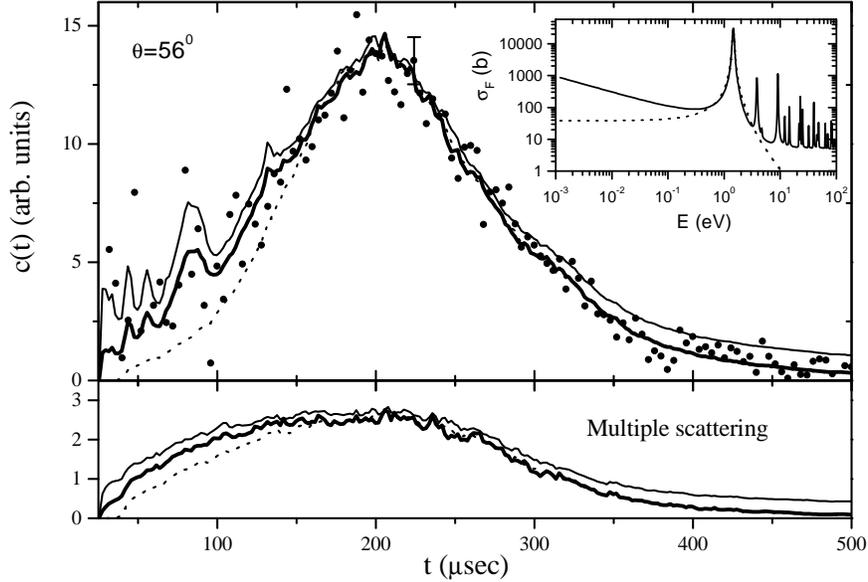}}
\caption{Main frame: Comparison of the Monte Carlo run for a 2mm thick 
polyethylene sample (thick line) with a black detector model (thin line) 
and with a Lorentzian shape for the filter total cross section (dotted line)
and the experimental data (dots).
Lower frame: detail of the multiple scattering components. Inset: Lorentzian
approach (dotted line) compared with the real total cross section for the 
indium filter (full line).}
\label{efflor}
\end{centering}
\end{figure}

In both cases commonly employed approaches are used. The results are
shown in Fig .  \ref{efflor} and in the inset the Lorentzian used to
represent the filter together with the exact cross section
\cite{Mcla}. We observe that both assumptions are inadequate and
affect both the single and the multiple scattering components. In the
case (a) the defect is manifested in an inaccurate description of the
long-times tail. On the other hand, in (b) we observe an incorrect
description in the short time region and the long-time tail, is mostly
unaffected.  It is worth to discuss the reason for both behaviors. The
long-times region of the observed profile is mainly composed by
emerging slow neutrons, which are absorbed by the filter according to
the above referred '1/v' behavior. In the case of a black detector (a)
our system is sensitive to those neutrons, while in (b) and our
detectors are covered with cadmium cylinders, whereby our detection
system is insensitive to such neutrons. Both behaviors are observed
in the calculated curves in Fig.~\ref{efflor}, while experimental data
only marginally illustrate the effect due to experimental errors.

\begin{figure}
\begin{centering}
\resizebox{0.8\textwidth}{!}{\includegraphics{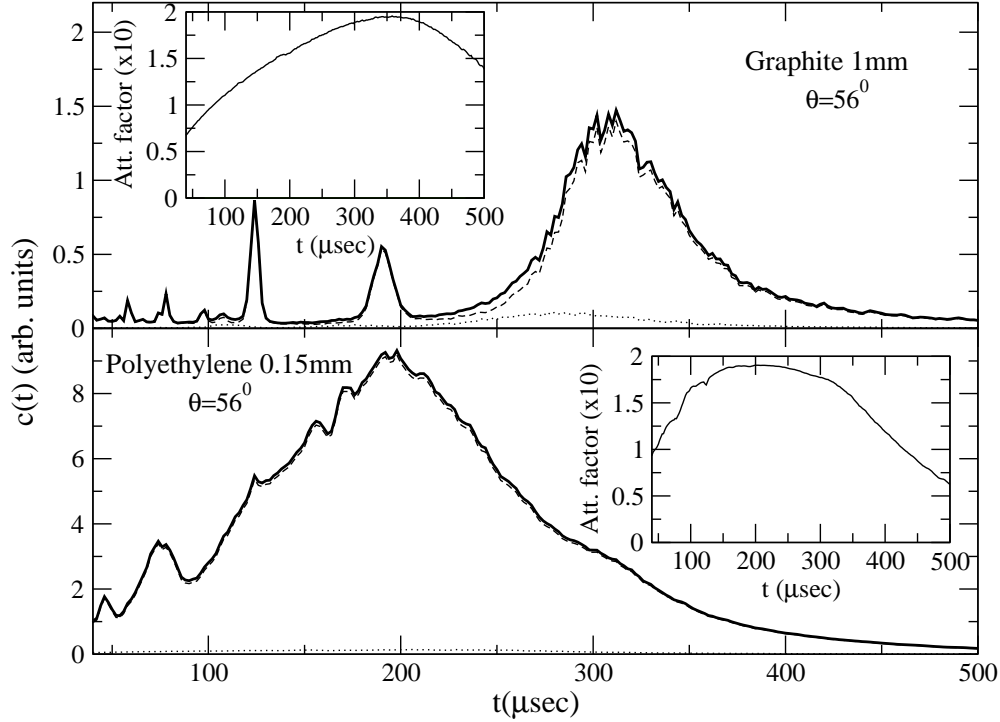}}
\caption{NCP for thin graphite and polyethylene samples chosen to approach the
  experimentalists' needs. The same notation as in Fig. \ref{graphite}
  applies. Insets: attenuation factors}
\label{realsize}
\end{centering}
\end{figure}

Finally, it is worth to mention that in common practice, the
experimentalist will choose an adequate sample size in order to
minimize multiple scattering effects, while keeping an acceptable
signal-to-noise ratio. To illustrate the multiple scattering and
attenuation effects on samples suitable to the experimentalist, we
present in Fig.~\ref{realsize} the results of our Monte Carlo program
for thin sample thicknesses of graphite and polyethylene of the same
diameters as presented in Figs.~\ref{graphite} and \ref{poly}. In the
upper frame we show our results for graphite (1 mm thick). We observe
that although the multiple scattering contribution is small, it is non
negligible, and it will have to be properly computed if accurate
values of the peak-shape parameters are to be obtained from the
experiment. Special attention must be payed to the attenuation factor,
that still has an appreciable variation over the range of times of
interest. The reason is that $H(\mathbf {k}_0,\mathbf {k})$ in Eq.
(\ref{factrans}) contains not only the attenuation in the sample
(negligible for a thin sample) but also the fraction of detected
neutrons (detector efficiency effects). In the case of polyethylene,
we show the results of a 0.15 mm thick sample. The multiple scattering
effect is barely visible, but the attenuation factor has also an
appreciable variation over the range of interest, thus affecting the
observed peak shape.

\section{Discussion and conclusions} 
 
Throughout this paper we examined different aspects that affect
multiple scattering and attenuation effects in DINS experiments. We
presented a Monte Carlo procedure that adequately describes Multiple
Scattering and attenuation processes in DINS experiments. To attain a
good agreement between the numerical simulations and the experimental
data, accurate descriptions of the incident neutron spectrum, the
detector efficiency as a function of the energy and the filter total
cross section were necessary. These considerations add up to those
stated in Refs. \cite{Blos,Blos2}, regarding the need of a good
description of the experimental setup and the inadequacy of the
convolution approximation in the description of the neutron Compton
profiles.  It must be remarked the importance of an accurate
description of the efficiency of the detector system. As such, we
understand the detecting setup that comprises the kind of detectors
employed, the geometry and the materials involved. For instance, the
detector system employed in this work is composed by $^3$He tubes
covered with cadmium, that result in the efficiency shown in Fig.
\ref{eff}. The importance of the knowledge of the detector efficiency
as a function of the energy can be understood in the light of the
recent work on the analysis of final energy distributions in DINS
experiments \cite{finalE}. As a result of this analysis it was
concluded that at every time of flight a distribution of final
energies is operative in the neutron Compton profiles, instead of a
single well-defined energy as is usually assumed in the customary data
processing procedures \cite{MayeMS}. As a consequence is is not
sufficient to assume a detector operative at a single energy (in which
case the knowledge of the efficiency function would be irrelevant),
but the behavior of the efficiency function is essential, as
confirmed by the results presented in Fig. \ref{efflor}.

The results presented on Multiple Scattering effects should be
considered when designing a DINS experiment. In the analyzed graphite
samples, the multiple scattering components have a significant
structure that affect the shapes of the neutron Compton profiles.  In
the case of polyethylene, multiple scattering contributes with a flat
background, but on the other hand the attenuation factor has a
significant variation over the time range of interest. This is a
consequence of the substantial variation of the efficiency function of
our detection system in the range of energies shown in Fig.
\ref{eff}.

From the analyzed examples, we conclude that a wise choice of the
sample thickness is still a valid rule, taking into account that a
good contrast between `filter-out' and `filter-in' positions is
required. Even if the sample sizes are adequately chosen, multiple
scattering and/or attenuation corrections will necessarily have to be
considered. The case of Fig.~\ref{realsize} is illustrative. Even if
the multiple scattering contribution is small, the correction due to
the attenuation factor can still be important (as shown in our case),
given that it also includes the detectors' efficiency correction
\cite{Daw1}, which in our case has an appreciable variation with the
energy. In general this result will depend on each particular
detection system. All the considered cases show the need to perform
accurate multiple scattering, attenuation and efficiency corrections.
To this end numerical simulations are the most adequate procedure.  In
this paper we presented a suitable correction tool, for which an
experimental benchmark with samples considerably affected by these
corrections was satisfactorily performed.

These corrections will normally have to be taken into account before
proceeding to the data analysis, {\it i.e.}  obtaining the kinetic
energy distributions of the atoms, peak areas, etc. Particular
importance will have the corrections in the analysis of lighter
nuclei. Monte Carlo simulations on heavier nuclei not shown in this
paper \cite{MayeMS} reveal that the multiple scattering component
tends to be located below the main peak and it is roughly proportional
to it, thus having negligible distortion effects on its shape. This
consideration must be brought together with those mentioned in Ref.
\cite{Blos} regarding the extreme care that must be exercised when
analyzing light nuclei with the DINS technique.

\section{Acknowledgements} 
 
We acknowledge Dr. R.E. Mayer for his collaboration during the
experiments.  We are especially grateful to L. Ca\-pa\-ra\-ro, M.
Schnee\-be\-li and P.  D'Avanzo for the technical support.  This work
was supported by ANPCyT (Argentina) under Project PICT No. 03-4122,
and CONICET (Project PEI 149/98).


\begin{thebibliography}{99} 
\bibitem{Hohe} P. C. Hohenberg and P.M. Platzmann, Phys. Rev.
  \textbf{152,} 198, (1966).
\bibitem{May1} J. Mayers, Phys. Rev. Lett. \textbf{71,} 1553, (1993).
\bibitem{LosAlamos} R. M. Brugger, A. D. Taylor, C. E. Olsen, J. A.
  Goldstone and A. K. Soper, Nucl. Instr. and Meth. {\bf 221}, 393
  (1984).
\bibitem{LosAlamos2} R. M. Brugger and P. A. Seeger, Nucl. Instr.  and
  Meth. {\bf A236}, 423 (1985).
\bibitem{Japon} H. Rauh and N. Watanabe, Nucl. Instr. and Meth. {\bf
    222}, 507 (1984).
\bibitem{Wang} Y. Wang and P. E. Sokol, Phys. Rev. Lett. {\bf 72},
  1040, (1994).
\bibitem{Dawn2m} J. Dawidowski, J. J. Blostein and J. R. Granada, in:
  M. R.  Johnson, G. J.  Kearley, H. G. B\"{u}ttner (Eds.) Neutron and
  Numerical Methods, American Institute of Physics, New York, (1999),
  p. 37
\bibitem{Blosmun}J. J. Blostein, J. Dawidowski, J. R. Granada and R. E.
  Mayer, Appl. Phys. A {\bf 74} [Suppl.], S157 (2002).
\bibitem{Blosjap} J. J. Blostein, J. Dawidowski, and J. R. Granada,
  Proceedings of the 15th. International Collaboration on Advanced
  Neutron Sources, J. Suzuki and S. Itoh (Eds.), Tsukuba, Japan, (2000),
  p. 689.
\bibitem{Sea1} V. F. Sears, Phys. Rev. B \textbf{30,} 44, (1984).
\bibitem{May2} J. Mayers, Phys. Rev. B \textbf{41,} 41, (1990).
\bibitem{Blos} J. J. Blostein, J. Dawidowski and J. R. Granada, Physica
  B \textbf{304,} 357, (2001).
\bibitem{Blos2} J. J. Blostein, J. Dawidowski and J. R. Granada,
  Physica B, {\bf 334}, 257 (2003).
\bibitem{PRB} J.J. Blostein, J.  Dawidowski and J.R. Granada, Phys. Rev.
  B {\bf 71}, 054105 (2005).
\bibitem{Daw1} J. Dawidowski, F. J. Bermejo and J. R. Granada, Phys.
  Rev. B \textbf{58,} 706, (1998).
\bibitem{Vine} G. M. Vineyard, Phys. Rev.\textbf{96,} 93, (1954).
\bibitem{Blec} I. A. Blech and R. L. Averbach, Phys. Rev. A
  \textbf{137,} 1113, (1965).
\bibitem{Sea2} V. F. Sears, Adv. Phys. \textbf{24,} 1, (1975).
\bibitem{Copl} J. R. D. Copley, Comp. Phys. Comm. \textbf{7,} 289,
  (1974); J. R. D. Copley, P. Verkerk, A. A. Van Well and H. Fredrikze,
  Comp.  Phys. Comm. \textbf{40,} 337, (1986).
\bibitem{Daw2} J. Dawidowski, J. R. Granada, R. E. Mayer, G. J. Cuello,
  V. H. Gillette and M.-C. Bellissent-Funel \textbf{203,} 116, (1994).
\bibitem{MayeMS} J. Mayers, A. L. Fielding and R. Senesi, Nucl. Instr.
  and Meth. A {\bf 481}, 454, (2002).
\bibitem{Fiel} A. L. Fielding and J. Mayers, Nucl. Instr. and Meth. A
  {\bf 480}, 680, (2002).
\bibitem{finalE}J. J. Blostein, J. Dawidowski and J. R. Granada,
  Nucl. Instr. and Meth. B (submitted, 2003).
\bibitem{Daw0} J. Dawidowski, G. J. Cuello and J. R. Granada,
  Nucl. Instr. and Meth B {\bf 82}, 459 (1993).
\bibitem{Gra2} J. R. Granada, Phys. Rev. B \textbf{31,} 4167, (1985).
\bibitem{Powl} J. G. Powles, Mol. Phys. \textbf{26,} 1352, (1976).
\bibitem{Bisc} F. G. Bischoff, M. L. Yeater and W. E. Moore, Nucl. Sci.
  Eng. \textbf{48,} 266, (1972).
\bibitem{Span} J. Spanier and E. Gelbard, \textit{Monte Carlo
    principles and neutron problems}, Addison Wesley, Reading, (1969).
\bibitem{Blat} J. M. Blatt and V. F. Weisskopff, \textit{Theoretical
    Nuclear Physycs}, Wiley, New York (1952), p. 470.
\bibitem{Mcla} V. McLane, C. L. Dunford and P. F. Rose, \textit{Neutron
    Cross Sections}, Vol. 2, Academic Press, New York, 1988, p. 411.
\bibitem{Ashc} N. W. Ashcroft and N. D. Mermin, \textit{Solid State
    Physics}, Saunders College Publishing, 1976.
\bibitem{Gra1} J. R. Granada, Z. Naturforsch. \textbf{39a,} 1160 ,
  (1984).
\bibitem{Daw3} J. R. Granada, J. Dawidowski, R. E. Mayer and V. H.
  Gillette, Nucl. Instr.  Meth. \textbf{A261,} 573, (1987).
\bibitem{Mugh} S. F. Mughabghab, M. Divadeenam and N. E. Holden, {\it
    Neutron Cross Sections}, Academic Press, Vol. 1, Part B, p. 79-1
  (1981).
\bibitem{Beck} K. H. Beckurts and K. Wirtz, {\it Neutron Physics},
  Springer, Berlin, 1964, p. 134.
\end {thebibliography}
\end{document}